\documentclass{ws-procs9x6}
\newcommand{\ud}{\mathrm{d}}  
\begin{document}
\thispagestyle{empty}                                                          
                                                                               
\begin{center}                                                                 
\begin{tabular}{p{130mm}}                                                      
                                                                               
\begin{center}                                                                 
{\bf\Large                                                                     
PATTERN FORMATION} \\                                                            
\vspace{5mm}                                                                   
                                
{\bf\Large IN WIGNER-LIKE EQUATIONS}\\
\vspace{5mm}

{\bf\Large VIA MULTIRESOLUTION}\\

\vspace{1cm}

{\bf\Large Antonina N. Fedorova, Michael G. Zeitlin}\\

\vspace{1cm}

{\bf\it
IPME RAS, St.~Petersburg,
V.O. Bolshoj pr., 61, 199178, Russia}\\
{\bf\large\it e-mail: zeitlin@math.ipme.ru}\\
{\bf\large\it e-mail: anton@math.ipme.ru}\\
{\bf\large\it http://www.ipme.ru/zeitlin.html}\\
{\bf\large\it http://www.ipme.nw.ru/zeitlin.html}
\end{center}

\vspace{1cm}

\abstracts{
We present the application of the variational-wavelet analysis to                  
the quasiclassical calculations of the solutions of Wigner/von Neumann/Moyal and related equations
corresponding
to the nonlinear (polynomial) dynamical problems. (Naive) deformation                     
quantization, the multiresolution 
representations and the variational approach are the key points. We        
construct the solutions via the multiscale expansions in the generalized          
coherent states or high-localized nonlinear eigenmodes in the base of the          
compactly supported wavelets and the wavelet packets. We demonstrate the appearance of (stable)
localized patterns (waveletons) and consider entanglement and decoherence as possible applications. 
}
\vspace{10mm}

\begin{center}
{\large Presented at Joint 28th ICFA Advanced Beam Dynamics  }\\
{\large \& Advanced \& Novel Accelerators Workshop on}\\
{\large QUANTUM ASPECTS OF BEAM PHYSICS } \\
{\large and Other Critical Issues of Beams in Physics and Astrophysics}\\
{\large January 7-11, 2003, Hiroshima University, Higashi-Hiroshima, Japan}
\end{center}
\end{tabular}
\end{center}
\newpage

\title{Pattern Formation in Wigner-like Equations via Multiresolution}
\author{A.~N. Fedorova and M.~G. Zeitlin\footnote{http://www.ipme.ru/zeitlin.html,
http://www.ipme.nw.ru/zeitlin.html}}
\address{IPME RAS, St.~Petersburg,\\
V.O. Bolshoj pr., 61, 199178, Russia\\
E-mail: zeitlin@math.ipme.ru, anton@math.ipme.ru}

\maketitle                                             

\abstracts{ 
We present the application of the variational-wavelet analysis to                  
the quasiclassical calculations of the solutions of Wigner/von Neumann/Moyal and related equations
corresponding
to the nonlinear (polynomial) dynamical problems. (Naive) deformation                     
quantization, the multiresolution 
representations and the variational approach are the key points. We        
construct the solutions via the multiscale expansions in the generalized          
coherent states or high-localized nonlinear eigenmodes in the base of the          
compactly supported wavelets and the wavelet packets. We demonstrate the appearance of (stable)
localized patterns (waveletons) and consider entanglement and decoherence as possible applications.    
}

\section{Wigner-like Equations}

In this paper we consider the  applications of 
a nu\-me\-ri\-cal\--\-ana\-ly\-ti\-cal technique based on local nonlinear harmonic analysis
(wavelet analysis, generalized coherent states analysis) 
to  the quasiclassical
calculations in nonlinear (polynomial) dynamical problems in the Wigner-Moyal approach.
The corresponding class of Hamiltonians has the form
\begin{eqnarray}
\hat{H}(\hat{p},\hat{q})=\frac{\hat{p}^2}{2m}+U(\hat{p},\hat{q}),
\end{eqnarray}
where $U(\hat{p}, \hat{q})$ is an arbitrary polynomial 
function on $\hat{p}$, $\hat{q}$, 
and plays the key role in many areas of physics [1], [2]. The particular cases, 
related to some physics models, are considered in [3]-[12].
Our goals are some attempt of classification and the explicit numerical-analytical constructions
of the existing quantum states in the class of models under consideration.
There is a hope on the understanding of relation between the structure of initial Hamiltonians and
the possible types of quantum states and the qualitative type of their behaviour.
Inside the full spectrum  there are at least three possibilities which are the most 
important from our point of view:
localized states, chaotic-like or/and entangled patterns, localized (stable) patterns 
(definitions can be found below).
All such states are interesting in the different areas of physics (e.g., [1], [2]) discussed below. 
Our starting point is the general point of view of a deformation quantization approach at least on
the naive Moyal/Weyl/Wigner level [1], [2].
The main point of such approach is based on ideas from [1], which allow  
to consider the algebras of quantum observables as the deformations
of commutative algebras of classical observables (functions). 
So, if we have as a model for classical dynamics the classical counterpart of 
Hamiltonian (1) and the Poisson manifold $M$ (or symplectic 
manifold or Lie coalgebra, etc) as the corresponding phase space, 
then for quantum calculations we need first of all to find
an associative (but non-commutative) star product 
 $*$ on the space of formal power series in $\hbar$ with
coefficients in the space of smooth functions on $M$ such that
\begin{eqnarray}
f * g =fg+\hbar\{f,g\}+\sum_{n\ge 2}\hbar^n B_n(f,g), 
\end{eqnarray}
where
$\{f,g\}$
is the Poisson brackets, $B_n$ are bidifferential operators.
Kontsevich gave the solution to this deformation problem in terms of the formal
power series via the sum over graphs and proved that for every Poisson manifold M there is a canonically
defined gauge equivalence class of star-products on M. Also there are the 
nonperturbative corrections to power 
series representation for $*$  [1]. In the naive calculations we may use the simple formal rule:
\begin{eqnarray}
* &\equiv&\exp \Big(\frac{i\hbar}{2}(\overleftarrow\partial_q\overrightarrow\partial_p-
   \overleftarrow\partial_p\overrightarrow\partial_q)\Big)
\end{eqnarray}
In this paper we consider the calculations of the Wigner functions
$W(p,q,t)$ (WF) corresponding
to the classical polynomial Hamiltonian $H(p,q,t)$ as the solution
of the Wigner-von Neumann equation [2]:
\begin{eqnarray}
i\hbar\frac{\partial}{\partial t}W = H * W - W * H
\end{eqnarray}
and related Wigner-like equations.
According to the Weyl transform, a quantum state (wave function or density operator $\rho$) corresponds
to the Wigner function, which is the analogue in some sense of classical phase-space distribution [2].
We consider the following form of differential equations for time-dependent WF, $W=W(p,q,t)$:
\begin{eqnarray}
W_t=\frac{2}{\hbar}\sin\Big[\frac{\hbar}{2}
(\partial^H_q\partial^W_p-\partial^H_p\partial^W_q)\Big]\cdot HW
\end{eqnarray}
which is a result of the Weyl transform of von Neumann equation
\begin{equation}
i\hbar\frac{\partial\rho}{\partial t}=[H,\rho]
\end{equation}
In our case (1) we have the following decomposition [2]
($U=U(q)$ in the following only for simplicity, but the case $U=U(p,q)$ can be
considered analogously):
\begin{equation}
\frac{\partial W}{\partial t}=T+U,
\end{equation}
where
\begin{equation}
U=\sum_{i=0}^\infty\frac{(i\hbar/2)^{2l}}{(2l+1)!}
\frac{d^{2l+1}U(q)}{dq^{2l+1}}\frac{\partial^{2l+1}}{\partial p^{2l+1}}W(p,q;t)
\end{equation}
\begin{equation}
T=-\frac{p}{m}\frac{\partial}{\partial q}W(p,q;t)
\end{equation}
Let $\{|E\rangle, E\}$ be the full set of discrete/continuous eigenfunctions (eigenvalues) 
\begin{equation}
H|E\rangle=E|E\rangle
\end{equation}
then we have the following representation for the Moyal function:
\begin{equation}
W_{E'',E'}(q,p)\equiv\frac{1}{2\pi\hbar}\int^\infty_{-\infty}
d\xi e^{-ip\xi/h}(q+\frac{1}{2}\xi|E''\rangle\langle E'|q-\frac{1}{2}\xi\rangle
\end{equation}
which is reduced to the standard WF in the case $E'=E''$:
$W_{E,E}(p,q)\equiv W(p,q)$.
As a result, the time independent Moyal function generates 
the time evolution of the WF.
The corresponding integral representation contains the initial value of the density operator 
$\rho(0)$ as a factor [2].
The Moyal function satisfies the following system of (pseudo)differential equations
\begin{eqnarray}
&&\Big[\frac{p^2}{2m}+U-\frac{\hbar^2}{8m}\frac{\partial^2}{\partial q^2}
+\sum^\infty_{i=1}\frac{(-1)^l(\hbar/2)^{2l}}{(2l)!}
\frac{d^{2l}U}{dq^{2l}}\frac{\partial^{2l}}{\partial p^{2l}}
\Big]
W_{E'',E'}\nonumber\\
&&=\frac{E'+E''}{2}W_{E'',E'}
\end{eqnarray}
\begin{eqnarray}
&&\Big[\frac{p}{m}\frac{\partial}{\partial q}-\frac{dU}{dq}\frac{\partial}{\partial p}
-\sum^\infty_{i=1}\frac{(-1)^l(\hbar/2)^{2l}}{(2l+1)!}
\frac{d^{2l+1}U}{dq^{2l+1}}\frac{\partial^{2l+1}}{\partial p^{2l+1}}
\Big]
W_{E'',E'}\nonumber\\
&&=\frac{i}{\hbar}(E''-E')W_{E'',E'}
\end{eqnarray}
really nonlocal/pseudodifferential for arbitrary Hamiltonians.
But in case of polynomial Hamiltonians (1)
we have only a finite number of terms in the corresponding series.
Also, in the stationary case after Weyl-Wigner mapping we have the following 
equation on WF in c-numbers [2]:
\begin{eqnarray}
&&\Big( \frac{p^2}{2m}+\frac{\hbar}{2i}\frac{p}{m}\frac{\partial}{\partial q}-
 \frac{\hbar^2}{8m}\frac{\partial^2}{\partial q^2}\Big)W(p,q)+\\
&& U\Big(q-\frac{\hbar}{2i}\frac{\partial}{\partial p}\Big)W(p,q)=\epsilon W(p,q)\nonumber
\end{eqnarray}
After expanding the potential $U$ into the Taylor 
series we have two real partial differential equations which correspond
to the mentioned before particular case of the Moyal equations  (12), (13).

Our approach, presented below, in some sense is motivated by the analysis of the following
standard simple model considered in [2].
Let us consider model of interaction of nonresonant atom with quantized electromagnetic field:
\begin{eqnarray}
\hat{H}=\frac{\hat{p}_x^2}{2m}+U(\hat{x}),\qquad
U(\hat{x})=U_0(z,t)g(\hat{x})\hat{a}^+\hat{a}
\end{eqnarray}
where potential $U$ 
depends on creation/annihilation operators and some polynomial on $\hat{x}$ 
operator function (or approximation)
$g(\hat{x})$.
It is possible to solve Schroedinger equation
\begin{eqnarray}
i\hbar\frac{\ud|\Psi>}{\ud t}=\hat{H}|\Psi>
\end{eqnarray}
by the simple ansatz 
\begin{eqnarray}
|\Psi(t)>=\sum_{-\infty}^{\infty}w_n\int\ud x|\Psi_n(x,t)|x>\otimes|n>
\end{eqnarray}
which leads to the hierarchy of analogous equations with potentials created by 
n-particle Fock subspaces
\begin{eqnarray}
i\hbar\frac{\partial\Psi_n(x,t)}{\partial t}=\Big\{\frac{\hat{p}_x^2}{2m}+
 U_0(t)g(x)n\Big\}\Psi_n(x,t)
\end{eqnarray}
where
$\Psi_n(x,t)$ is the probability amplitude of finding the atom at 
the time $t$ at the position $x$ and the field in the $n$ Fock state.
Instead of this, we may apply the Wigner approach starting with proper full density matrix
\begin{eqnarray}
&&\hat{\rho}=|\Psi(t)><\Psi(t)|=\\
&&\sum_{n',n''}w_{n'}w^*_{n''}\int\ud x'\int\ud x''
\Psi_{n'}(x',t)\Psi^*_{n''}(x'',t)|x'><x''|\otimes|n'><n''|\nonumber
\end{eqnarray}
Standard reduction gives pure atomic density matrix
\begin{eqnarray}
&&\hat{\rho}_a\equiv\int^{\infty}_{n=0}<n|\hat{\rho}|n>=\\
&&\sum|w_n|^2
\int\ud x'\int\ud x''\Psi_n(x',t)\Psi^*_n(x'',t)|x'><x''|\nonumber
\end{eqnarray}
Then we have incoherent superposition 
\begin{equation}
W(x,p,t)=\sum^{\infty}_{n=0}|w_n|^2W_n(x,p,t)
\end{equation}
of the atomic Wigner functions
\begin{equation}
W_n(x,p,t)\equiv\frac{1}{2\pi\hbar}\int\ud\xi{\rm exp}\Big(-\frac{i}{\hbar}p\xi\Big)
\Psi^*_n(x-\frac{1}{2}\xi,t)\Psi_n(x+\frac{1}{2}\xi,t)
\end{equation}
corresponding to the atom motion in the potential $U_n(x)$ 
(which is not more than polynomial in $x$) generated by $n$-level Fock state. 
They are solutions of proper Wigner equations:
\begin{eqnarray}
\frac{\partial W_n}{\partial t}=-\frac{p}{m}\frac{\partial W_n}{\partial x}+
\sum^{\infty}_{\ell=0}\frac{(-1)^\ell(\hbar/2)^{2\ell}}{(2\ell+1)!}
\frac{\partial^{2\ell+1}U_n(x)}{\partial x^{2\ell+1}}
\frac{\partial^{2\ell+1}W_n}{\partial p^{2\ell+1}}
\end{eqnarray}

In the following section we'll generalize this construction and we are interested in
description of entanglement in the Wigner formalism.

The next example describes the decoherence process.
Let we have collective and environment subsystem with their own Hilbert spaces 
\begin{equation}
\mathcal{H}=\mathcal{H}_c\otimes\mathcal{H}_e
\end{equation}
Relevant dynamics are described by three parts including interaction
\begin{equation}
H=H_c\otimes I_e+I_c\otimes H_e+H_{int}
\end{equation}

For analysis, we can choose Lindblad master equation [2]
\begin{eqnarray}
\dot{\rho}=\frac{1}{i\hbar}[H,\rho]-\sum_n\gamma_n(L^+_nL_n\rho+
\rho L^+_nL_n-2L_n\rho L^+_n)
\end{eqnarray}
which preserves the positivity of density matrix and it is Markovian
but it is not general form of exact master equation.
Other choice is Wigner transform of master equation [2] and it is more preferable for us
\begin{eqnarray}
&&\dot{W}=\{H,W\}_{PB}+\\
&&\sum_{n\geq 1}\frac{\hbar^{2n}(-1)^n}{2^{2n}(2n+1)!}
\partial^{2n+1}_x U(x)\partial_p^{2n+1}W(x,p)+
2\gamma\partial_p pW+D\partial^2_pW\nonumber
\end{eqnarray}

In the next section we consider the variation-wavelet approach for the solution of all
these Wigner-like equations (6)-(9), (12), (14), (23), (26), (27) for the case of an 
arbitrary polynomial $U(q, p)$, which corresponds to a finite number 
of terms in the series from (8), (12), (13), (14), (23), (27) 
or to proper finite order of $\hbar$.
Our approach is based on the extension of our variational-wavelet 
approach [3]-[12].
Wavelet analysis is some set of mathematical methods, which gives the possibility to
work with  well-localized bases in functional spaces and gives maximum sparse
forms for the general type of operators (differential, integral, pseudodifferential) in such bases.
These bases are the natural generalization of standard coherent, squeezed, thermal squeezed states [2],
which correspond to quadratical systems (pure linear dynamics) with Gaussian Wigner functions.
Because the affine
group of translations and dilations (or more general group, which acts on the space of solutions) 
is inside the approach
(in wavelet case), this
method resembles the action of a microscope. 
We have a contribution to
the final result from each scale of resolution from the whole underlying 
infinite scale of spaces. 
Our main goals are an attempt of classification and construction 
of possible nontrivial states
in the system under consideration.
We are interested in the following states: localized, entangled patterns,
localized (stable) patterns. 
We start from the corresponding definitions 
(at this stage these definitions have only
qualitative character).

1. By localized state (localized mode) 
we mean the corresponding (particular) solution of the system under 
consideration which is localized in maximally small region of the phase space.

2. By chaotic/entangled pattern we mean some solution (or asymptotics of solution) 
of the system under consideration
which has equidistribution of energy spectrum in a full domain of definition. 

3. By localized pattern (waveleton) 
we mean (asymptotically) stable solution localized in 
relatively small region of the whole phase space (or a domain of definition). 
In this case all energy is distributed during some time (sufficiently large) 
between few localized modes (from point 1) only.

Numerical calculations
explicitly demonstrate the quantum interference of
generalized coherent states, pattern formation from localized eigenmodes and 
the appearance of (stable) localized patterns (waveletons).

\section{Variational Multiscale Representation}

We obtain our multiscale/multiresolution representations for solutions of Wig\-ner-like equations
via a variational-wavelet approach. 
We represent the solutions as 
decomposition into modes related to the hidden underlying set of scales [13]:
{\setlength\arraycolsep{0pt} 
\begin{eqnarray}
&&W(t,q,p)=\displaystyle\bigoplus^\infty_{i=i_c}\delta^iW(t,q,p)
\end{eqnarray}}
where value $i_c$ corresponds to the coarsest level of resolution
$c$ or to the internal scale with the number $c$ in the full multiresolution decomposition
of underlying functional space ($L^2$, e.g.) corresponding to problem under consideration:
\begin{equation}
V_c\subset V_{c+1}\subset V_{c+2}\subset\dots
\end{equation}
and $p=(p_1,p_2,...), q=(q_1,q_2,...), x_i=(p_1,q_1,....,p_i,q_i)$ 
are coordinates in phase space.
In the following we may consider as fixed as variable 
numbers of particles. The second case corresponds to quantum statistical ensemble 
(via ``wignerization'' procedure) and will be considered in details elsewhere [12].

We introduce the Fock-like space structure

\begin{eqnarray}
H=\sum^{\infty}_{n=0}\bigoplus H^n
\end{eqnarray}
for the set of n-particle wave functions (states):
\begin{eqnarray}
W=\{W_0,W_1(x_1;t),W_2(x_1,x_2;t),\dots,
W_N(x_1,\dots,x_N;t),\dots\},
\end{eqnarray}
where
$W_p(x_1,\dots, x_p;t)\in H^p$,
$H^0=C,\quad H^p=L^2(R^{6p})$ (or any different proper functional space), $W\in$
$H^\infty=H^0\oplus H^1\oplus\dots\oplus H^p\oplus\dots$
with the natural Fock space like norm 
(guaranteeing the positivity of the full measure):
\begin{eqnarray}
(W,W)=W^2_0+\sum_{i}\int W^2_i(x_1,\dots,x_i;t)\prod^i_{\ell=1}\mu_\ell.
\end{eqnarray}
First of all we consider $W=W(t)$ as a function of time only,
$W\in L^2(R)$, via
multiresolution decomposition which naturally and efficiently introduces 
the infinite sequence of the underlying hidden scales [13].
We have the contribution to
the final result from each scale of resolution from the whole
infinite scale of spaces. 
We consider a multiresolution decomposition of $L^2(R)$
(of course, we may consider any different and proper for some particular case functional space)
which is a sequence of increasing closed subspaces $V_j\in L^2(R)$ 
(subspaces for 
modes with fixed dilation value):
\begin{equation}
...V_{-2}\subset V_{-1}\subset V_0\subset V_{1}\subset V_{2}\subset ...
\end{equation}
The closed subspace
$V_j (j\in {\bf Z})$ corresponds to  the level $j$ of resolution, 
or to the scale j
and satisfies
the following properties:
let $W_j$ be the orthonormal complement of $V_j$ with respect to $V_{j+1}$: 
$
V_{j+1}=V_j\bigoplus W_j.
$
Then we have the following decomposition:
\begin{eqnarray}
\{W(t)\}=\bigoplus_{-\infty<j<\infty} W_j \qquad {\rm or} \qquad
\{W(t)\}=\overline{V_0\displaystyle\bigoplus^\infty_{j=0} W_j},
\end{eqnarray}
in case when $V_0$ is the coarsest scale of resolution.
The subgroup of translations generates a basis for the fixed scale number:
$
{\rm span}_{k\in Z}\{2^{j/2}\Psi(2^jt-k)\}=W_j.
$
The whole basis is generated by action of the full affine group:
\begin{eqnarray}
{\rm span}_{k\in Z, j\in Z}\{2^{j/2}\Psi(2^jt-k)\}=
{\rm span}_{k,j\in Z}\{\Psi_{j,k}\}
=\{W(t)\}.
\end{eqnarray}
Let the sequence $\{V_j^t\}, V_j^t\subset L^2(R)$ 
correspond to multiresolution analysis on the time axis, 
$\{V_j^{x_i}\}$ correspond to multiresolution analysis for coordinate $x_i$,
then
$
V_j^{n+1}=V^{x_1}_j\otimes\dots\otimes V^{x_n}_j\otimes  V^t_j
$
corresponds to the multiresolution analysis for 
the $n$-particle function 
$W_n(x_1,\dots,x_n;t)$.
E.g., for $n=2$:$\qquad V^2_0=\{f:f(x_1,x_2)=$
{\setlength\arraycolsep{0mm}
$
\sum_{k_1,k_2}a_{k_1,k_2}\phi^2(x_1-k_1,x_2-k_2),\ 
a_{k_1,k_2}\in\ell^2(Z^2)\},
$}
where 
$
\phi^2(x_1,x_2)=\phi^1(x_1)\phi^2(x_2)=\phi^1\otimes\phi^2(x_1,x_2),
$
and $\phi^i(x_i)\equiv\phi(x_i)$ form a multiresolution basis corresponding to
$\{V_j^{x_i}\}$.
If $\{\phi^1(x_1-\ell)\},\ \ell\in Z$ form an orthonormal set, then 
$\phi^2(x_1-k_1, x_2-k_2)$ form an orthonormal basis for $V^2_0$.
So, the action of the affine group generates multiresolution representation of
$L^2(R^2)$. After introducing the detail spaces $W^2_j$, we have, e.g. 
$
V^2_1=V^2_0\oplus W^2_0.
$
Then the 
3-component basis for $W^2_0$ is generated by 
the translations of three functions 
\begin{eqnarray}
\Psi^2_1=\phi^1(x_1)\otimes\Psi^2(x_2),\ \Psi^2_2=\Psi^1(x_1)\otimes\phi^2(x_2), \ 
\Psi^2_3=\Psi^1(x_1)\otimes\Psi^2(x_2).\nonumber
\end{eqnarray}
Also, we may use the rectangle lattice of scales and one-dimensional wavelet
decomposition:
$$
f(x_1,x_2)=\sum_{i,\ell;j,k}\langle f,\Psi_{i,\ell}\otimes\Psi_{j,k}\rangle
\Psi_{j,\ell}\otimes\Psi_{j,k}(x_1,x_2),
$$
where the basis functions $\Psi_{i,\ell}\otimes\Psi_{j,k}$ depend on
two scales $2^{-i}$ and $2^{-j}$.
After constructing the multidimensional basis 
we may apply one of the variational procedures from [3]-[12].
We obtain our multiscale\-/mul\-ti\-re\-so\-lu\-ti\-on 
representations (formulae (40) below) 
via the variational wavelet approach for 
the following formal representation of the systems from the Section 1 
(or its approximations).
 
Let $L$ be an arbitrary (non)li\-ne\-ar dif\-fe\-ren\-ti\-al\-/\-in\-teg\-ral operator 
 with matrix dimension $d$
(finite or infinite), 
which acts on some set of functions
from $L^2(\Omega^{\otimes^n})$:  
$\quad\Psi\equiv\Psi(t,x_1,x_2,\dots)=\Big(\Psi^1(t,x_1,x_2,\dots), \dots$,
$\Psi^d(t,x_1,x_2,\dots)\Big)$,
 $\quad x_i\in\Omega\subset{\bf R}^6$, $n$ is the number of particles:
\begin{equation}
L\Psi\equiv L(Q,t,x_i)\Psi(t,x_i)=0,
\end{equation}
where
\begin{eqnarray}
&&Q\equiv Q_{d_0,d_1,d_2,\dots}(t,x_1,x_2,\dots,\partial /\partial t,\partial /\partial x_1,
\partial /\partial x_2,\dots,\int \mu_k)=\nonumber\\
&&\sum_{i_0,i_1,i_2,\dots=1}^{d_0,d_1,d_2,\dots}
q_{i_0i_1i_2\dots}(t,x_1,x_2,\dots)
\Big(\frac{\partial}{\partial t}\Big)^{i_0}\Big(\frac{\partial}{\partial x_1}\Big)^{i_1}
\Big(\frac{\partial}{\partial x_2}\Big)^{i_2}\dots\int\mu_k. 
\end{eqnarray}
Let us consider now the $N$ mode approximation for the solution as 
the following ansatz:
\begin{equation}
\Psi^N(t,x_1,x_2,\dots)=\sum^N_{i_0,i_1,i_2,\dots=1}a_{i_0i_1i_2\dots} A_{i_0}\otimes 
B_{i_1}\otimes C_{i_2}\dots(t,x_1,x_2,\dots).
\end{equation}
We shall determine the expansion coefficients from the following conditions
(different related variational approaches are considered in 
[3]-[12]:
\begin{equation}
\ell^N_{k_0,k_1,k_2,\dots}\equiv
\int(L\Psi^N)A_{k_0}(t)B_{k_1}(x_1)C_{k_2}(x_2)\ud t\ud x_1\ud x_2\dots=0.
\end{equation}
Thus, we have exactly $dN^n$ algebraical equations for  $dN^n$ unknowns 
$a_{i_0,i_1,\dots}$.
This variational approach reduces the initial problem 
to the problem of solution 
of functional equations at the first stage and 
some algebraical problems at the second.
 We consider the multiresolution expansion as the second main part of our 
construction. 
So, the solution is parametrized by the solutions of two sets of 
reduced algebraical
problems, one is linear or nonlinear
(depending on the structure of the operator $L$) and the rest are linear
problems related to the computation of the coefficients of the algebraic equations (39).
These coefficients can be found  by some wavelet methods
by using the
compactly supported wavelet basis functions for the expansions (38).
As a result the solution of the equations from Section 1 has the 
following mul\-ti\-sca\-le or mul\-ti\-re\-so\-lu\-ti\-on decomposition via 
nonlinear high\--lo\-ca\-li\-zed eigenmodes 
\begin{eqnarray}
&&W(t,x_1,x_2,\dots)=
\sum_{(i,j)\in Z^2}a_{ij}U^i\otimes V^j(t,x_1,x_2,\dots),\nonumber\\
&&V^j(t)=
V_N^{j,slow}(t)+\sum_{l\geq N}V^j_l(\omega_lt), \quad \omega_l\sim 2^l, \\
&&U^i(x_s)=
U_M^{i,slow}(x_s)+\sum_{m\geq M}U^i_m(k^{s}_mx_s), \quad k^{s}_m\sim 2^m,
 \nonumber
\end{eqnarray}
which corresponds to the full multiresolution expansion in all underlying time/space 
scales.
The formulae (40) give the expansion into a slow part
and fast oscillating parts for arbitrary $N, M$.  So, we may move
from the coarse scales of resolution to the 
finest ones for obtaining more detailed information about the dynamical process.
In this way one obtains contributions to the full solution
from each scale of resolution or each time/space scale or from each nonlinear eigenmode.
It should be noted that such representations 
give the best possible localization
properties in the corresponding (phase)space/time coordinates. 
Formulae (40) do not use perturbation
techniques or linearization procedures.
Numerical calculations are based on compactly supported
wavelets and related wavelet families [13] and on evaluation of the 
accuracy on 
the level $N$ of the corresponding cut-off of the full system 
regarding norm (32):
\begin{equation}
\|W^{N+1}-W^{N}\|\leq\varepsilon.
\end{equation}

So, by using wavelet bases with their best (phase) space/time      
localization  properties we can describe localized (coherent) structures in      
quantum systems with complicated behaviour.
The modeling demonstrates the appearance of different (stable) pattern formation from
high-localized coherent structures or chaotic behaviour.
Our (nonlinear) eigenmodes are more realistic for the modelling of 
nonlinear classical/quantum dynamical process  than the corresponding linear gaussian-like
coherent states. Here we mention only the best convergence properties of the expansions 
based on wavelet packets, which  realize the minimal Shannon entropy property
and the exponential control of convergence of expansions like (40) based on the norm (32).
Fig. 1 shows the high-localized eigenmode contribution to the WF, while Fig. 2, 3 give 
the representations for the full solutions, constructed
from the first 6 eigenmodes (6 levels in formula (40)), and demonstrate the stable localized 
pattern formation (waveleton) and complex 
chaotic-like behaviour. Fig. 3 corresponds to (possible) result of superselection
(einselection) [2] after decoherence process started from Fig. 2 or Fig. 4.
Fig. 5 and Fig. 6 demonstrate time steps during appearance of entangled states.
It should be noted that
we can control the type of behaviour on the level of the reduced algebraical system (39)
[12]. 

\section*{Acknowledgements}

We are very grateful to Prof. Pisin Chen (SLAC) for invaluable encouragement, support
and patience.

\newpage
\begin{figure}[ht]
\centerline{\epsfxsize=3.6in\epsfbox{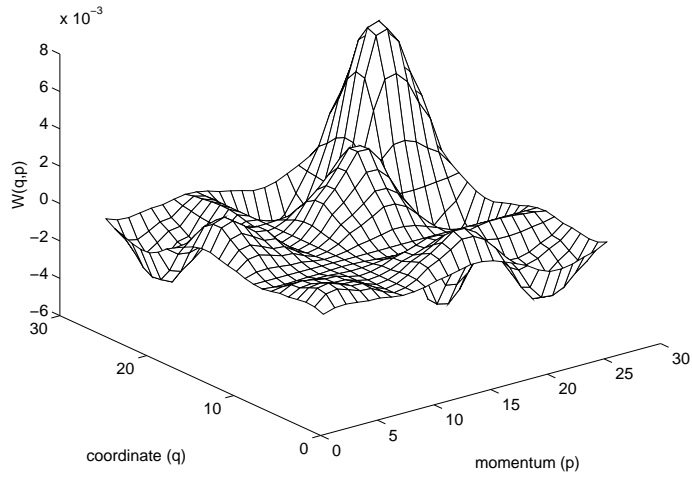}}
\caption{Localized mode contribution to Wigner function.}
\end{figure}
\begin{figure}[ht]
\centerline{\epsfxsize=3.6in\epsfbox{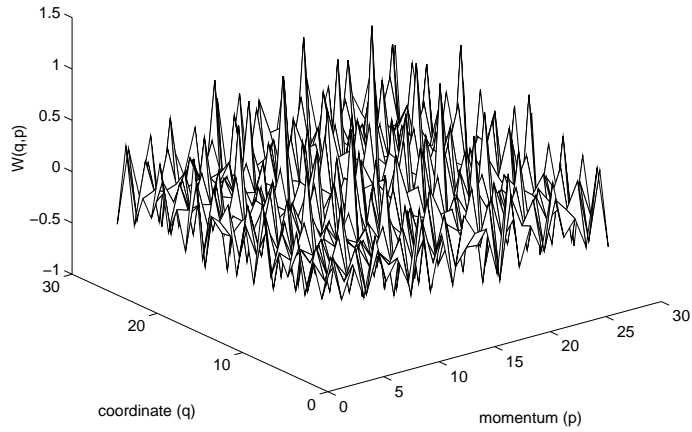}}
\caption{Chaotic-like Wig\-ner function.}
\end{figure}
\begin{figure}[ht]
\centerline{\epsfxsize=3.9in\epsfbox{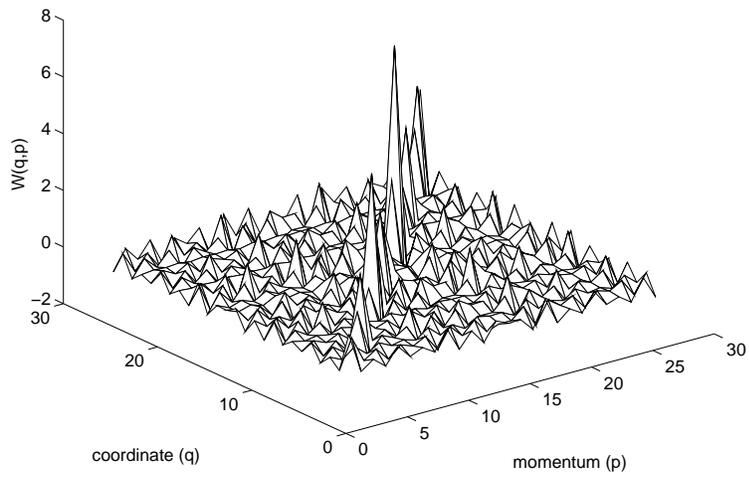}}
\caption{Localized pattern-like (waveleton) Wigner function.}
\end{figure}
\begin{figure}[ht]
\centerline{\epsfxsize=3.6in\epsfbox{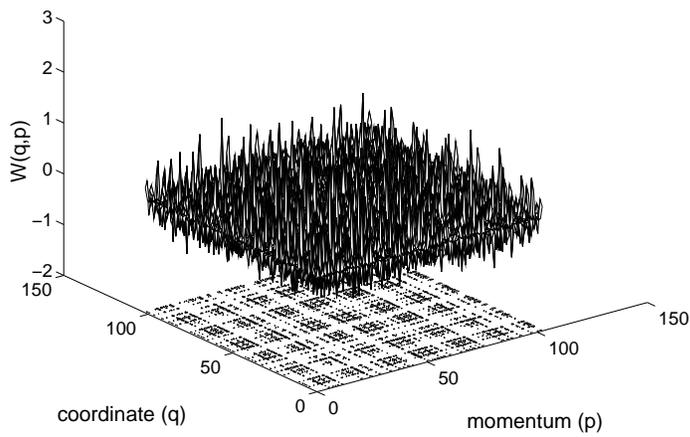}}
\caption{Entangled-like Wigner function.}
\end{figure}
\begin{figure}[ht]
\centerline{\epsfxsize=3.6in\epsfbox{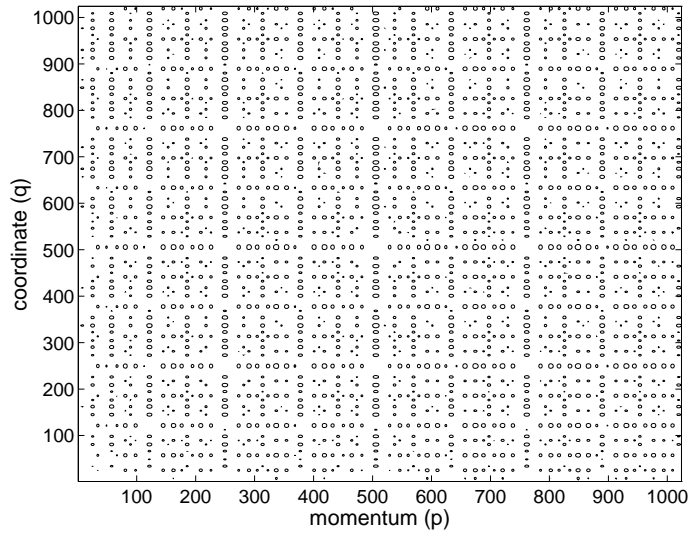}}
\caption{Section of the coarse level approximation for Wigner function.}
\end{figure}
\begin{figure}[ht]
\centerline{\epsfxsize=3.9in\epsfbox{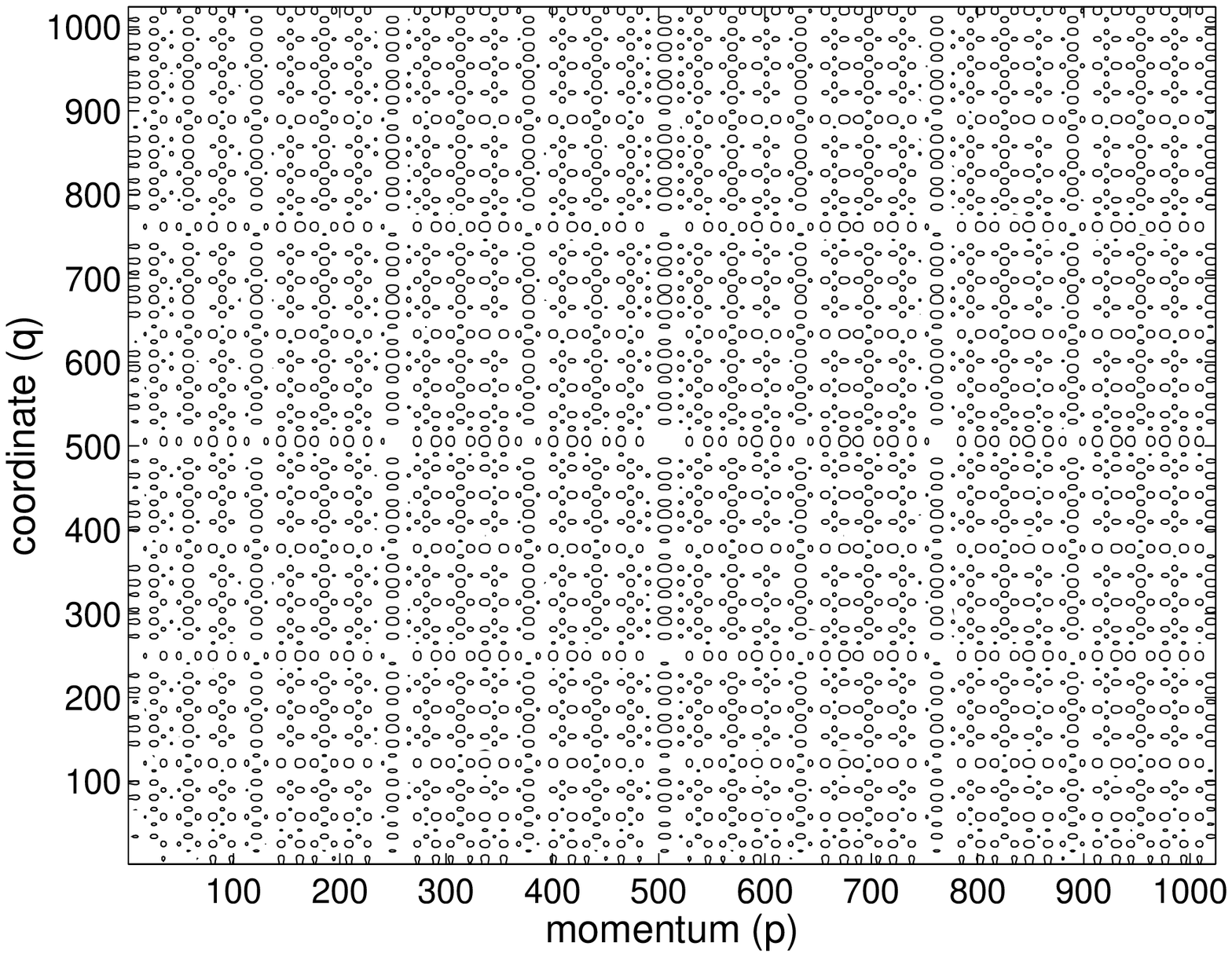}}
\caption{Section of the finest level approximation for Wigner function.}
\end{figure}


\begin{thebibliography}{13}

\bibitem{1}                                                                    
D. Sternheimer, Los Alamos preprint, math.QA/9809056, 
T. Curtright, T. Uematsu, C. Zachos, Los Alamos preprint: hep-th/0011137.

\bibitem{2} 
W. P. Schleich, Quantum Optics in Phase Space, Wiley, 2000,
W. Zurek, Los Alamos preprint: quant-ph/0105127.

\bibitem{3}
A.N. Fedorova, M.G. Zeitlin, 
 {\it Math. and Comp. in Simulation}, {\bf 46}, 527, 1998.

\bibitem{4}
A.N. Fedorova, M.G. Zeitlin,
{\it New Applications of Nonlinear and Chaotic Dynamics in Mechanics}, Ed. F. Moon, 31, 101
Klu\-wer,  1998.

\bibitem{5}
A.N. Fedorova, M.G. Zeitlin,
{\bf CP405}, 87, American Institute of Physics, 1997.
Los Alamos preprint: phy\-sics\-/\-9710035.

\bibitem{6}
A.N. Fedorova, M.G. Zeitlin and Z.~Parsa,    
{\bf CP468}, 48, 69, American Institute of Physics, 1999.
Los Alamos preprints: physics/990262, 9902063.


\bibitem{7}
A.N. Fedorova, M.G. Zeitlin,
The Physics of High Brightness Beams, Ed. J. Rosenzweig, 235, World Scientific, 2001. 
Los Alamos preprint: physics/0003095.


\bibitem{8}
A.N. Fedorova, M.G. Zeitlin, Quantum Aspects of Beam Physics, 
Ed. P. Chen, 527, 539, World Scientific, 2002; arXiv preprints: phy\-sics\-/0101006, 0101007.

\bibitem{9}
A.N. Fedorova, M.G. Zeitlin, 
Proceedings EPAC2002, pp. 1323, 1344, 1434, 1482, 1595, EPS-IGA/CERN, 2002,
arXiv preprints: physics/0206049, 0206050, 0206051, 0206052, 0206053.

\bibitem{10}
A.N. Fedorova, M.G. Zeitlin,
Proceedings in Applied Mathematics and Mechanics, Volume 1, Issue 1,
399, 432, Wiley-VCH, 2002, arXiv preprints: nlin.PS/0206024, physics/0206054.

\bibitem{11} 
A.N. Fedorova, M.G. Zeitlin,
Nuclear Inst. and Methods in Physics Research, A,
vol 502/2-3 657, 660, 2003,
arXiv preprints: quant-ph/0212166, physics/0212115.

\bibitem{12}
A.N. Fedorova, M.G. Zeitlin,
Progress in Nonequilibrium Green's Functions II,
Ed. M. Bonitz and D. Semkat, World Scientific, 481, 2003,
arXiv preprint: physics/0212066 and in press.

\bibitem{13}
Y. Meyer, {\it Wavelets and Operators}, Cambridge Univ. Press, 1990.

\end{thebibliography}
\end{document}